\documentstyle[12pt,a4]{article}
\def\Pom{{I\!\!P}}
\def\Reg{{I\!\!R}}
\begin{document}
%
% \draft
%
\date{}
\title{ 
Deviations from the superposition model in a Dual
Parton Model  
with formation zone cascade in both projectile and
target nuclei  
}
\author{G.\ Battistoni\\[2mm]
{\it INFN, Sezione di Milano, I--20133 Milano, Italy }\\[5mm]
 C.\ Forti\\[2mm]
{\it INFN, Laboratori Nazionali di Frascati,I--00044 Frascati, Italy }\\[5mm]
 J.\ Ranft\\[2mm]
{\it Departamento de F\'\i sica de Part\'\i culas,}\\
{\it Universidade de
Santiago de Compostela,}\\
{\it E--15706 Santiago de Compostela, Spain}\\[5mm]
 S.\ Roesler\\[2mm]
{\it Universit\"at Siegen, Fachbereich Physik,D--57068 Siegen, Germany}\\[5mm]
}

\maketitle

\begin{abstract}
A Dual Parton Model with a formation zone intranuclear cascade 
in the spectators of the projectile and target nuclei is
studied. The hadrons produced in the formation zone cascade
contribute to Feynman--$x_F$ and lab--$x$ distributions in the
fragmentation regions of the target and projectile nuclei.
We discuss the consequences of this model in the secondary cosmic
ray production, by analyzing
the calculated spectrum weighted moments for pion and kaon
production.
We show that the proposed model leads to significant differences
with respect to a simple superposition model, where the nucleus--nucleus
collision is replaced by a few corresponding nucleon-nucleus collisions.
\end{abstract}

%\begin{center}
%{\footnotesize
%PACS number(s): 96.40.Pq, 96.40.Tv, 13.85.Hd, 13.85.Ni, 13.85.Tp } \\
%{\footnotesize
%Keywords: High Energy Cosmic Rays, Nucleus--Nucleus Collisions, 
%Dual Parton Model, Glauber model}
%\end{center}
\vspace*{-22cm}
\hfill Preprint US--FT / 29--96 \\
\vspace*{-21cm}
\hfill revised version, October 1996

\clearpage
%-------------------- Introduction (Sect. 1) ---------------------------
%
\section {Introduction}
\noindent
Nucleus-nucleus collisions are of great importance for the understanding
of the cosmic ray cascades in the atmosphere. Models for sampling hadron
production events in nucleus--nucleus and hadron--nucleus collisions are
needed for the simulation of the development of cosmic ray showers in
atmosphere. In some models used for the simulation of cosmic ray
cascades, examples are HEMAS \cite{HEMAS} and SIBYLL
\cite{SIBYLL}, the correct treatment of nucleus--nucleus
collisions is replaced by a simpler superposition model
\footnote{it must be noticed that SIBYLL takes however into account
the correct differences in the interaction heights as coming
from the nucleus--nucleus cross sections.}, where
the nucleus--nucleus collision is replaced by the
corresponding nucleon--nucleus collisions. Here we study a
model where the nucleus--nucleus collisions are treated in a
more detailed way.

Soft and hard multiple  interactions between nucleons of both nuclei
dominate the hadron production in most of the kinematic 
region covered by
the interaction and are well described in the framework of the Dual
Parton Model (DPM)~\cite{Capella94a}. 
They were extensively studied with
the Monte Carlo (MC) implementation of this model:
 {\sc dpmjet-ii}~\cite{Ranft94c}.
This  event generator
 was already applied for sampling cosmic ray cascades
\cite{Ranft94c,DPMBFR94,dpmcharm}. 
However, when dealing with particle or fragment production in the 
forward or backward fragmentation regions a detailed description of 
intranuclear cascade processes and of nuclear disintegration is
important.
A formation zone intranuclear cascade in the target nucleus was
therefore considered since a few years
\cite{Ranft88a,Ranft89a,Moehring91,Ranft94c}.
A formation zone intranuclear cascade (FZIC) model in both the target 
as well as the projectile nucleus, the calculation
of nuclear excitation energies, models for nuclear evaporation, 
high energy fission and break-up of light nuclei were discussed for
hadron--nucleus and nucleus--nucleus collisions
in~\cite{Ferrari95a,Ferrari96a}. It was shown that these MC implementations
for hadron--nucleus and nucleus--nucleus 
interactions describe successfully the basic features 
of target and projectile associated particle production.

The model was compared in
 \cite{Ferrari95a,Ferrari96a} mostly to data
for asymmetric collisions in which projectile and target masses are
different. It seems that the performance of the
model in these collisions is rather good. Asymmetric collisions are the 
most common ones also in cosmic ray cascades.

Here we will study this model with then aim to find
differences with respect to the superposition models mentioned above.

In Section 2 we summarize briefly the main steps of sampling 
hadron--nucleus and nucleus--nucleus 
interactions within the event generator  
{\sc dpmjet-ii}. Furthermore,
we summarize the basic ideas of the FZIC
model. In Section 3 we calculate Feynman--$x_F$ distributions of
pions produced in hadron--nucleus and nucleus--nucleus
collisions.
In Section 4 we study spectrum--weighted moments  and energy
fractions carried away by some kinds of secondaries  and we compare them
with the expectations from superposition models. It is pointed
out that the DPM differs from the superposition model.
 Finally, in Section 5 we summarize our results.
%
%-------------------- Sect. 2 ------------------------------------------
%

%
\section{The two-component DPM for hadron-nucleus
            and nucleus-nucleus collisions}

\subsection{The DPMJET--II event generator}
The two-component DPM and its MC realizations for
hadron--hadron, hadron--nucleus and nucleus--nucleus collisions
have been discussed in detail 
in~\cite{Aurenche92a,Bopp94a,Ranft94c}. 
Therefore, we briefly summarize the main steps leading to the 
multiparticle state, which is the starting point for the intranuclear 
cascade.

The MC model for hadron-nucleus and nucleus-nucleus
interactions starts from an impulse approximation for the
nucleons of the interacting nuclei. The spatial initial configuration,
i.e. the positions of the nucleons in space-time in the rest system of
the corresponding nucleus, is sampled from standard density distributions. 
For energies above 3-5~GeV/nucleon
the collision proceeds via $\nu$ elementary interactions
between $\nu_p$ and $\nu_t$ nucleons from the projectile and target,
respectively. The values $\nu, \nu_p,$ and $\nu_t$ are sampled according 
to Glauber's multiple scattering formalism using the MC algorithm
of~\cite{Shmakov89}. The particle production 
is well described by the two-component DPM which is applied as in
hadron-hadron interactions~\cite{Aurenche92a,Bopp94a}.
As a result, a system of chains
connecting partons of the nucleons involved in the scattering process is
formed. The chains are hadronized applying the 
{\footnotesize JETSET} model~\cite{Sjostrand86,Sjostrand87a}. 
The hadrons may then cause
intranuclear cascade processes, which are treated by the FZIC 
model~\cite{Ranft88a}, an extension of the
intranuclear cascade model of ref.~\cite{Bertini63,Bertini69}.
At energies below 3-5~GeV/nucleon the FZIC model is able by 
itself to describe reasonably the inelastic nuclear collisions.

\subsection{The formation zone cascade in target and projectile
nuclei}

In the following we summarize the main ideas of the FZIC 
model for hadron-nucleus  and
nucleus--nucleus interactions
\cite{Ranft88a,Ranft89a,Moehring91,Ranft94c,Ferrari95a,Ferrari96a}.
The physical picture explaining the absence of the intranuclear cascade
at high energies is the concept of the formation zone~\cite{Stodolski75}. 
It has been introduced in analogy to the 
Landau-Pomeranchuk~\cite{Landau53} effect, which explains
the observation that electrons passing through high density
materials become more
penetrating at high energies. For the formation zone of an electron 
with 4-momentum $p$ and energy $E$ upon
radiation of a photon with 4-momentum $k$ one obtains
\begin{equation}
\label{tauelec}
\tau=\frac{E}{k\cdot p}=\frac{E}{m}\frac{1}{\omega_e},
\end{equation}
where $\omega_e$ is the frequency of the photon in the rest frame of the
electron and $E/m$ is the time dilatation factor from the electron rest
frame to the laboratory.
Within the quark model, the states being
formed in the primary nucleon-nucleon interaction can be understood as
consisting of valence quarks only (i.e without the full system of sea
quarks, antiquarks and gluons) and have therefore a reduced probability
for hadronic interactions inside the nucleus~\cite{Ranft88a}. 
The formation zone concept can be translated to hadron production as
follows~\cite{Ranft89a}. 
%%
%% revision 16-01-96
We consider the formation zone cascade in the rest system of the target
nucleus (laboratory system) or in the rest system of the
projectile nucleus.
Denoting the 4-momenta of the projectile hadron $p_p$ and of the secondary
hadron $p_s$ 
%%
%% revision 16-01-96
%in the laboratory frame
%%
with
\begin{equation}
p_p=(E_p,0,0,\sqrt{E_p^2-m_p^2}), \qquad 
p_s=(E_s,\vec{p}_{s\perp},\sqrt{E_s^2-m_s^2-\vec{p}_{s\perp}^2})
\end{equation}
and replacing in Eq.~(\ref{tauelec}) the electron momentum by $p_p$ and
the photon momentum by $p_s$,
%Denoting the 4-momentum of the projectile hadron with
%$p_p=(E_p,0,0,\sqrt{E_p^2-m_p^2})$, that of the secondary hadron with
%$p_s=(E_s,\vec{p}_{s\perp},\sqrt{E_s^2-m_s^2-\vec{p}_{s\perp}^2})$, and
%using Eq.~(\ref{tauelec}) 
%the hadron formation zone in the laboratory frame can be written as
the hadron formation time is, for $E_p\gg m_p$ :
\begin{equation}
\tau_{\mbox{\scriptsize Lab}}=\frac{2E_s}{(m_px)^2+m_s^2+p_{s\perp}^2}, 
\qquad
x=\frac{E_s}{E_p}.
\end{equation}
The term $(m_px)^2$ can be
neglected for most of the produced secondaries, so one can approximate
\begin{equation}
\tau_{\mbox{\scriptsize Lab}}\approx \gamma_s\tau_s, \qquad 
\gamma_s=\frac{E_s}{m_s}.
\end{equation}
We define an average formation time to create a complete
hadronic state $\tau_s$ in the rest system 
of the secondary hadron $s$
~\cite{Ranft88a,Ranft89a}:
\begin{equation}
\label{deffortim}
\tau_s=\tau_0\frac{m_s^2}{m_s^2+p_{s\perp}^2}.
\end{equation}
where $\tau_0$ is a free parameter, which has to be determined by 
comparing particle production within the model to experimental data. 
Typical values are in the range from 1~fm/$c$ to 10~fm/$c$.
Here we use  $\tau_0=4.5$~fm/$c$. This value differs somewhat
from the values reported in ref.
\cite{Ferrari95a,Ferrari96a}. The reason for this is an updated
treatment of nuclear evaporation with respect to the version of the code
used in \cite{Ferrari95a,Ferrari96a}.
For each secondary we sample a formation time $\tau$ from an exponential
distribution~\cite{Bialas84} with an average value as given in
Eq.~(\ref{deffortim}). 
As it was described in~\cite{Moehring91}, 
in the MC model  the full
space-time history of the collision is known.

After having assigned a formation time to a secondary, its spatial
coordinates in the rest system of both nuclei are known and we start
with considering an intranuclear cascade step in one (randomly chosen) 
of the spectators.

Due to relativistic time dilatation, those secondaries which
have sufficiently high energies in the rest frame of the considered nucleus 
are
formed mostly outside of the spectator part of this nucleus, whereas
those with lower energies are formed inside. The latter may penetrate the
spectator and initiate intranuclear cascade processes. 
Elastic and inelastic interactions with spectator nucleons are treated 
using the MC-model {\sc hadrin}~\cite{Haenssgen86}. This code
is based on measured cross sections and interaction channels up to a
laboratory momentum of 5~GeV. We apply {\sc hadrin} to
hadron-nucleon interactions up to 9~GeV and neglect those at higher
energies. Reinteractions beyond 5~GeV occur much less frequently than 
reinteractions below 5~GeV and a more detailed treatment would not
change the results discussed in this paper. Furthermore, we take into account
absorption of low-energy mesons and antiprotons by interactions with
two-nucleon systems (for pion absorption we use the cross
sections as given by Ritchie~\cite{Ritchie83}) 
and Pauli's principle~\cite{Moehring91}. 
In case no interaction is possible in the 
considered spectator, we proceed with sampling a cascade step in another
spectator. 

For secondaries produced in intranuclear cascade processes we 
apply the same formalism, i.e. a formation time is sampled, the secondary is
transported to the end of the formation zone and reinteractions are
treated if they are possible. 
Due to these intranuclear cascade processes, nucleons are knocked 
out of the residual spectator nuclei if their energy is high enough to
escape from the nuclear potential.

\section{Feynman--$x_F$ distributions in hadron--nucleus and
nucleus--nucleus collisions}

The  so-called ``cascade particles'' are the particles which are 
produced or knocked--out off the spectator nucleus
by the FZIC. Target associated
cascade particles are experimentally studied (mostly in
emulsion experiments) and are often called ``grey'' prongs. 

In the papers \cite{Ferrari95a,Ferrari96a} most of  the
available data on grey particle production in high energy 
hadron--nucleus
and nucleus--nucleus collisions are compared to the model we use.
 Reasonable agreement is
found as far as the average multiplicities of grey prongs, their
multiplicity and angular distribution, and their correlations 
with the number of fast produced particles are concerned.
There are two kinds of
particles contributing to the grey prongs: protons and charged
pions. The emulsion experiments are not able to differentiate
between such protons and charged pions.  There is however one
heavy ion experiment (WA80 
\cite{Albrecht93,Albrecht93a}) which presents only protons in
the energy region of the grey prongs in their multiplicity
distributions. We can only draw some rather indirect
support for the ability of our model to describe the pions from
the intranuclear cascade. This support comes from the fact that 
the model agrees as
well to the emulsion data on grey prongs as to the WA80 data on
protons alone. Roughly 20\% of the grey prongs (defined
with Lorentz parameter 0.23 $< \beta <$ 0.70) in 
the model are charged pions\cite{Ferrari96a}.

The Feynman--$x_F$ distributions are the most suitable way to
present the particle production in the target and projectile
fragmentation regions.  Unfortunately, most experiments, which
measure grey prongs are  not able to measure the particle
momenta and identity and are therefore not able to present
Feynman--$x_F$ distributions. We define Feynman--$x_F$ in h--A and A--A
collisions with the
longitudinal momentum in the hadron-nucleon (or nucleon-nucleon)
 cms $p^{cms}_z$ :
\begin{equation}
x_F = \frac{p^{cms}_z}{|p^{cms}_{z,max}|}
\end{equation}
where we use a $p^{cms}_{z,max}$ calculated disregarding Fermi
momenta.  The Feynman--$x_F$ distribution $f(x_F)$ is defined as
follows
\begin{equation}
f(x_F) = x_F \frac{dN}{dx_F}.
\end{equation}

For cosmic ray calculations we are mainly interested in the
contributions of pions from the intranuclear cascade to the
Feynman--$x_F$ distributions in the fragmentation region of the
projectile nucleus, since particles from target fragmentation
are very slow in the lab--frame. However, experimental data about
the target fragmentation are very important for the tuning of the
interaction model.
The changes in Feynman--$x_F$ distributions from  p--p to p--nucleus 
collisions in the proton
fragmentation region have been measured (and have been compared
to DPMJET--II in \cite{Ranft94c}), but we are not aware of any
such measurements in the fragmentation region of the nucleus.
Therefore, we have to rely on the model, but we stress that it
would be highly desirable to measure Feynman--$x_F$ distributions
in the fragmentation region of target or projectile nuclei.

In Fig. \ref{pairxfpip1e3n} we present the Feynman--$x_F$
distribution of $\pi^+$ mesons in p--air collisions 
at a lab--energy of 500 TeV .
There are two plots, one for the full model with formation
zone intranuclear cascade in the target nucleus,
and one without this cascade. We also
compare the distributions with the one in p--p collisions at
the same energy. In the target fragmentation region, at $x_F$
values between -0.3 and -1, we find significant differences in
the distributions due to the pions produced by the formation
zone cascade in the target nucleus.  The difference between the two
plots is not due to a large number of particles. 
If we define ``grey'' the particles with Lorentz--$\beta$ of 
0.23 $ < \beta < $ 0.70 
in the lab--frame,  we find in the calculation 0.06 grey $\pi^+$ per 
interaction, on average.
However, not all charged pions from the formation zone cascade belong 
to the grey particles.

Next, we turn to nucleus--air collisions and calculate again
Feynman--$x_F$ distributions.  It is already enough to look at
Fig.\ref{pairxfpip1e3n} to understand, qualitatively, what
happens in nucleus-nucleus collisions. Fig.\ref{pairxfpip1e3n}
gives the Feynman $x_F$ distribution  in N-p collisions if we
simply exchange $x_F$ by $-x_F$. Therefore, we expect that the
FZIC will lead to significant changes of the distributions in the
fragmentation regions of nuclear projectiles in nucleus-proton
and nucleus--nucleus collisions.

In Figs. \ref{heairxfpip1e3n} to \ref{feairxfpip1e3n} 
we present the Feynman--$x_F$
distribution of $\pi^+$ mesons 
at a lab--energy of 500 TeV per projectile nucleon in He--air,
O--air,  and Fe--air collisions. The model, as shown
in \cite{Ranft94c}, exhibits a rather good Feynman--scaling
behaviour, so that these distributions look rather
similar also at other different energies. 
Fig.\ref{heairxfpip1e3n} is included mainly to
demonstrate that for a light nucleus, like He, the formation zone
cascade makes really no difference and can be neglected.
There are two plots on each figure,
one for the full model with formation
zone intranuclear cascade in the target as well as in the
projectile nucleus, and one without this cascade. We also compare the 
distributions with the corresponding p--air collisions at
the same energy and with the formation zone cascade. 
In the target fragmentation region, at $x_F$
values between -0.3 and -1 we find in each plot nearly the same 
differences in
the distributions due to the pions produced by the formation
zone cascade in the target nucleus. Such differences are also
visible in the projectile fragmentation region at positive $x_F$
values, but here those differences, which were not significant in
the case of He--air collisions, are found to rise with the mass number of
the projectile nucleus. Clearly, the formation zone cascade in
the spectators of the Fe nucleus produces more pions than the
one in the O--spectators.

\section{ Spectrum weighted moments, energy fractions
and comparison with
superposition models}

Following for instance the basic discussion of Ref.~\cite{gaistext},
we introduce a variable $x_{lab}$ similarly to Feynman--$x_F$, but this
time in the lab--frame :

\begin{equation}
x_{lab} = \frac{E_i}{E_0},
\end{equation}

$E_i$ is the lab--energy of a secondary particle $i$ and $E_0$ is the 
lab--energy of the projectile in a h--A collision (or the energy per 
projectile nucleon in a A--A collision).
We introduce $x_{lab}$ distributions $F(x_{lab})$ :

\begin{equation}
F_i(x_{lab}) = x_{lab}\frac{dN_i}{dx_{lab}}.
\end{equation}

We note
that the Feynman--$x_F$ distribution at positive $x_F$ in the
projectile fragmentation region is a very good approximation to
the $x_{lab}$ distribution. Therefore, the Feynman $x_F$
distributions given in the last Section give also a rather good
picture for  $x_{lab}$ distributions.

The cosmic ray spectrum weighted moments in A--B collisions 
are defined as moments of the
$F(x_{lab})$
\begin{equation}
Z^{A-B}_i = \int^{A_A}_0 (x_{lab})^{\gamma -1}
F^{A-B}_i(x_{lab})dx_{lab}.
\end{equation}
Here $-\gamma \simeq$ --1.7 is the power of the integral cosmic
ray energy spectrum.

The spectrum weighted moments for nucleon--air collisions,
as discussed in 
 Ref.~\cite{gaistext},
 determine the
uncorrelated fluxes of energetic particles in the atmosphere.
This result has been obtained for cosmic ray cascades
initiated by a hadron primary.
There, the spectrum weighted moments (which
change only slightly with the primary energy) are the same for
all generations of the cascade.  We can not apply this result
directly for our situation. 
A cosmic ray cascade initialized by
primary nuclei becomes after the first two to three generations
also a cascade with only hadrons participating. 
There are no theoretical arguments
for the relevance of the spectrum
weighted moments of nucleus--nucleus collisions for the cosmic ray cascade
as a whole. These moments should be 
relevant only for the first generations.
However, we might use the spectrum weighted moments
and energy fractions just as any other moment of the
Feynman--$x_F$ distribution with the aim to point
out differences between models. 
In particular, we use these moments to stress that the
superposition model for nucleus--nucleus collisions is in
reality only a modest approximation to the correct treatment of
nucleus--nucleus collisions.

We also introduce the energy fraction $K^{A-B}_i$.
In h--A collisions that is the
fraction of primary energy carried by secondaries of type
$i$; in A--B collisions it is this energy fraction multiplied
with the mass number $A$ of the projectile nucleus
\begin{equation}
K^{A-B}_i = \int^{A_A}_0 
F^{A-B}_i(x_{lab})dx_{lab},
\end{equation}
where $A_A$ is the mass number of the projectile nucleus A.
After recalling the concept of superposition model, 
in the following we shall compare the calculation of spectrum weighted 
moments
with our model in different conditions, pointing out the differences
with respect to a simple superposition model.

\subsection{The superposition model}
 Generally, we call superposition model the approximation 
in which  in the cosmic ray cascade the collision of a nucleus A,
with total energy E, against a target B, is treated as the superposition
of $A_A$ independent nucleon-B collisions, each nucleon 
having an energy E/$A_A$.
This model is based  on the hypothesis that, 
when the energy per nucleon of the 
 projectile is much larger than the single nucleon binding energy, 
 the $A_A$ nucleons will interact incoherently. 
   In the context of cosmic ray simulations, for a primary of total 
 energy E and mass number $A_A$ ($A_A$$>$1), 
 the cascade generated is equivalent 
 to the total effect of $A_A$ showers initiated by 
 $A_A$ independent nucleons 
 of energy E/$A_A$. For example, the cascade for a primary iron nucleus of 
 560 TeV would be simulated as the superposition of 26 proton + 30 neutron 
 initiated showers, each of 10 TeV. Examples of application of the 
 superposition model to cosmic ray calculations
can be found in Ref.~\cite{cosmicres1,nim85}. 

Here, in the comparison with DPMJET--II we will apply a less general
concept of superposition model which we call restricted
superposition model, making reference to
a single nucleus--nucleus interaction. In a minimum bias 
A--B collision,
not all $A_A$ projectile nucleons interact. The number of
interacting projectile nucleons is only $\nu^{A-B}_p$. For the
collisions considered  the average values of 
$\nu^{A-B}_p$ are always given in Tables
1 to 4. Obviously, in order to make a comparison with DPMJET--II, we
should use a restricted superposition model where only $\nu^{A-B}_p$
projectile nucleons interact.
In this restricted  superposition model we have  
for instance 
$Z_{\pi}^{sup A-B} = \nu^{A-B}_p Z_{\pi}^{p-B}$ 
(for the same energy/nucleon of the
projectile).

\subsection{The Glauber model in nucleus--nucleus collisions
and the restricted superposition model}

The Glauber model for nucleus--nucleus collisions
\cite{Glauber59} is used by most models for nucleus--nucleus and
nucleon--nucleus collisions. DPMJET uses the implementation 
of ref.~\cite{Shmakov89}.   This model has
properties different from those of the restricted 
superposition model. In order to understand that, we recall the
main formulae for inelastic nucleus--nucleus collisions. From
these expressions we understand that the Glauber model cannot be
reduced just to nuclear geometry.

We start with the scattering amplitude of two nuclei with mass
numbers A and B in the impact parameter representation
\cite{Franco68,Czyz69,Kofoed69}
\begin{equation}
F({\bf b})_{A-B}= <\psi^f_A \psi^f_B|1-
\prod_{j=1}^A\prod_{k=1}^{B}[1-\chi 
({\bf b}-{\bf s}_j+{\bf \tau}_k)]|\psi^i_B
\psi^i_A>
\end{equation}
where ${\bf b}$ is the impact parameter 
vector and ${\bf s}_j$ and ${\bf \tau}_k$
are the coordinates of the nucleons with respect to the centers
of the nuclei A and B , respectively, in the impact parameter
plane. The $\psi^i_{A,B}$ and $\psi^f_{A,B}$ are the initial
and final state wave functions of nuclei A and B, while $\chi({\bf b})$ is
the dimensionless 
elastic nucleon--nucleon scattering amplitude in the impact
parameter representation. The (Gaussian) 
parametrization of $\chi({\bf b})$
used in DPMJET--II was described in detail in \cite{Ranft94c}.

Furthermore, we use a simple assumption for the squares of the
state wave functions
\begin{equation}
|\psi_A|^2=\prod_{j=1}^A \rho_A({\bf s}_j,z_j),~~~~~~~~~~
|\psi_B|^2=\prod_{k=1}^B \rho_B({\bf \tau}_k,\xi_k)
\end{equation}
where $\rho_A$ and $\rho_B$ are the one particle densities of
the nuclei A and B.

From this we get the total A--B cross section
\begin{equation}
\sigma^{tot}_{A-B}= 2 Re \int d^2{\bf b}F({\bf b})_{A-B}.
\end{equation}
The inelastic A--B cross section is determined by the profile
function $\Gamma({\bf b})$
\begin{equation}
\sigma^{inel}_{A-B}=\int d^2{\bf b}
\Gamma({\bf b})
\end{equation}
 $\Gamma({\bf b})$ is defined as follows
\begin{equation}
\Gamma({\bf b})=\int\{1-\prod_{i=1}^A\prod_{j=1}^B(1-p_{ij})\}
\{\prod_{i=1}^A\rho_A({\bf r}_i)d^3{\bf r}_i\}
\{\prod_{j=1}^B\rho_B({\bf t}_j)d^3{\bf t}_j\}
\end{equation}
where 
\begin{equation}
p_{ij}=
 \chi({\bf b}-{\bf s}_i+{\bf \tau}_j)+\chi^*({\bf b}-{\bf s}_i+{\bf \tau}_j)
-\chi({\bf b}-{\bf s}_i+{\bf \tau}_j)\chi^*({\bf b}-{\bf s}_i+{\bf \tau}_j).
\end{equation}
This profile function, which contains more than the nuclear
geometry, is also used to sample in the Monte Carlo calculation
the impact parameters of the inelastic A--B collisions.

Sampling the impact parameters of nucleon--nucleus and nucleus--nucleus
collisions simply from geometry, will
certainly lead to a restricted superposition model.
Unfortunately, it is somewhat cumbersome to see, without the actual
numerical calculations, that the profile functions according to
the Glauber model for nucleon--nucleus
collisions differ sufficiently from the profile functions for
nucleus--nucleus collisions to prevent the restricted 
superposition model
to give identical results as the full model.
As an example related to
the case discussed in the next subsection, we show in Fig.\ref{profilef}
the profile functions for p--Nitrogen He--Nitrogen and
Fe--Nitrogen as function of $b$ = $|{\bf b}|$.  For p--Nitrogen
collisions,  the profile function 
never becomes completely absorptive ($\Gamma(b)$ = 1), even in the center 
($b$=0),
while in Fe--Nitrogen collisions $\Gamma(b)$ =1 is reached for $b <$ 5 fm.

\subsection{Calculation results, DPMJET--II without FZIC,
eva\-po\-ra\-tion and determination of the residual nuclei }

In Table~\ref{table1} and \ref{table2} we present spectrum
weighted moments and energy fractions 
as calculated in DPMJET--II in the model without
 FZIC, evaporation and consequent determination of the residual nuclei.
We consider p--air, He--air, O--air
and Fe--air collisions at laboratory energies per nucleon of 50
GeV, 500 TeV, and $5\cdot 10^{6}$ TeV, the latter energy 
($5\cdot 10^{18}$~eV)
is the absolute highest energy at which DPMJET--II is valid, at
present. We also give the average number of projectile nucleons 
 $\nu_p$ participating in
minimum bias collisions.

In Tables~\ref{table1} and \ref{table2} we present a further
comparison, at the two upper
energies, of the spectrum
weighted moments of pions and kaons and the energy fractions of
pions, kaons and baryons minus antibaryons according to the model,
 with the moments and energy fractions
obtained from the restricted 
superposition model. The baryons are here only
the secondary baryons from the collision. Spectator baryons
from the projectile or target nucleus do not contribute to the
energy fraction $K^{A-B}_{b-\bar b}$. Since the energy fraction
of the newly produced baryons (produced in $b \bar b$ pairs) is
equal to the energy fraction of the produced antibaryons,
the energy fraction $K^{A-B}_{b-\bar b}$ gives the energy fraction
retained by the baryons participating in the collision.
In the restricted superposition model the spectrum weighted moments and
energy fractions are given by the simple expressions
\begin{equation}
Z^{sup A-B}_i =  
\nu^{A-B}_p Z^{N-B}_i , 
\end{equation}
\begin{equation}
K^{sup A-B}_i =  
\nu^{A-B}_p K^{N-B}_i , 
\end{equation}
where N--B refers to a nucleon nucleus collision.

We know, without any calculation, that the Dual Parton Model for
nucleus--nucleus collisions differs from the restricted superposition
model. There are at least three reasons for these differences:
\begin{enumerate}

\item Properties of the Glauber cascade: 
The deviations of the Glauber model from the restricted 
superposition model
were already discussed in the previous subsection.
In Table \ref{table3}
we give for each collision considered the average numbers of
collisions according to the Glauber cascade. $\nu$, $\nu_p$ and
$\nu_t$ were already explained above.
In nucleon--nucleus collisions we have always
 $\nu^{N-B}$ = $\nu^{N-B}_t$ and
 $\nu^{N-B}_p$ = 1. 
In A--B collisions 
we have $\nu^{A-B}$, $\nu^{A-B}_t$ and $\nu^{A-B}_p$ 
all different from one.
%In nucleon--nucleus collisions the projectile nucleon has $\nu$
%= $\nu_t$ collisions with target nucleons. 
In the A--B collisions,
a projectile nucleon interacts, 
in average, $N^{A-B}_p$ = $\nu^{A-B} / \nu^{A-B}_p$
times. As seen from Table \ref{table3},  $N^{A-B}_p$ is always
different from
$\nu^{N-B}_t$ at the same energy,  $N^{A-B}_p$ being smaller than
the corresponding $\nu^{N-B}_t$. The average nucleon--nucleus
collision in an A--B collision has different properties compared
to an average
 nucleon--nucleus
collision. This is valid for any model constructed on the basis
of the Glauber cascade.

\item  
 We turn to  the DPM with  Glauber cascade, and let us for
 simplicity discuss the chain structure disregarding the
 multiple soft chains and multiple minijets in each
 nucleon--nucleon interaction, which are in
 addition present in each elementary Glauber collision.
We construct in nucleon--nucleus collisions one pair of
 valence--valence chains and $\nu^{N-B}_t$ -- 1 pairs of sea--valence
 chains. In the restricted superposition model for the A--B collision 
 we obtain $\nu^{A-B}_p$
 pairs of valence--valence chains and $\nu^{A-B}_p (\nu^{N-B}_t - 1)$
 pairs of sea--valence chains.

In A--B collisions 
we have $\nu^{A-B}$, $\nu^{A-B}_t$ and $\nu^{A-B}_p$ 
all different from one and in
Fe-Air collisions we have usually $ \nu^{A-B} > \nu^{A-B}_p >
\nu^{A-B}_t$. In this situation we might form $\nu^{A-B}_t$
valence--valence chain pairs, $\nu^{A-B}_p - \nu^{A-B}_t$ valence--sea
chain pairs and $\nu^{A-B} - \nu^{A-B}_p$ sea--sea chain pairs (this
is the simplest possibility, in actual Monte Carlo models like
DPMJET--II the chain ends available are connected in a random way).
We know already that only the number of valence--valence 
 chain pairs in the restricted superposition model and in the
correct model correspond to each other, the number of
sea--valence chain pairs will be different. Furthermore,  there
are never any sea--sea chains in the restricted superposition model.  

\item
The single diffractive cross sections in hadron--nucleus
collisions are well studied \cite{Roesler93,Ranft94b} and
included in the model. The cross sections for single
diffractive nucleus--nucleus collisions are straightforward to
calculate within the DPM, and they will probably have a size
similar to that obtained in
hadron--nucleus collisions. The  O--air or
Fe-air total cross sections are three to five times larger than
 the p--air total cross section. Therefore, the fraction of single
diffractive nucleus--nucleus  collisions will come
out to be a few times  smaller than in hadron--nucleus
collisions. For these reasons, and because of the lack of experimental
data on single diffractive nucleus--nucleus cross sections, this
component is not included at present in the DPMJET--II event
generator (however, the inclusion of this part would probably not 
change the situation).
Now, in the case of diffractive p--air collisions, where
the projectile proton leaves the collision only slightly
deflected and in which hadrons are produced only in the nucleus
fragmentation region,  nearly no contribution to the
moments of pions and kaons is achieved. Therefore, also this
effect is expected to lead to larger moments of charged pions
and kaons in the correct DPM with respect to the restricted 
superposition model. 

\end{enumerate}

Looking at 
Tables~\ref{table1} and \ref{table2} we find 
 our expectations fulfilled. The spectrum weighted moments  of
 pions and kaons according to DPMJET-II are 10 to 20 \% larger than
 the ones according to the restricted 
 superposition model. Also the energy
 fractions carried by pions and kaons are for He--air and O--air
 about 5 to 10 \% larger than in the restricted superposition model. In
 Fe--air these energy fractions behave in a more complicated
 way. In the Fe-air reaction, it would be better to define the
 restricted 
 superposition model in the Fe rest frame with the air nuclei as
 projectiles. At the same time, as expected, 
 the energy fractions carried by the participating nucleons are
 for Fe--air and O--air 
 in DPMJET--II  always 
 smaller than in the restricted superposition model, i.e. more
 energy is used for particle creation. Here the numbers obtained
 in He--air are more difficult to understand.
 We have
to remember that all numbers in the Tables are the result of Monte Carlo
calculations with typically 5\% statistical error. 

In summary, we may conclude that for the model without FZIC
we find an agreement  with
the restricted superposition model within approximately 10 to 20 \%.

\subsection{Calculation results, the full DPMJET--II model 
}

In Table~\ref{table3} and \ref{table4} we present spectrum
weighted moments and energy fractions 
as calculated in the full DPMJET--II  model.
We consider again p--air, He--air, O--air
 and Fe--air collisions at laboratory energies per nucleon of 50
GeV, 500 TeV, and $5\cdot 10^{6}$ TeV. 
We also give the number of Glauber collisions $\nu$
between $\nu_p$ projectile and $\nu_t$ target nucleons in
minimum bias collisions in Table~\ref{table3}.

In Tables~\ref{table3} and \ref{table4} we have a further comparison,
at two energies, of the spectrum
weighted moments of pions and kaons and of the energy fractions of
charged pions, charged kaons, baryons minus antibaryons  and of
nuclear fragments and residual nuclei, 
 with the moments obtained from the restricted superposition model. 
Here we have to use slightly different expressions for the
energy fractions in the restricted superposition model. The
reason is that all spectators in the full DPMJET--II appear among
the final state particles, either as evaporation protons or
neutrons or as nuclear fragments and residual nuclei. We are not
able to define nuclear fragments and residual nuclei in the
restricted superposition model. In this respect the features of
DPMJET--II are again different from those of
the restricted superposition model. We use the
expressions for moments and energy fractions 
for charged pions and kaons in the same form as given in the
last subsection.
For the energy fraction of $b- \bar b$ we use:
\begin{equation}
K^{sup A-B}_{b - \bar b} =  
\nu^{A-B}_p K^{N-B}_{b - \bar b} +(A_A - \nu^{A-B}_p)\cdot 1 . 
\end{equation}
Since we are not able to define an energy fraction $K_{n.f.}$
for nuclear fragments, we should compare the
$K^{supA-B}_{b-\bar b}$ in Table 4 with the sum of
$K^{A-B}_{b-\bar b}$ and $K^{A-B}_{n.f.}$.

The differences  between the spectrum weighted moments of the
pions in the restricted model and in the full model in 
Tables \ref{table1} and \ref{table3}
are easy to understand looking at the previously
presented Feynman--$x_F$ distributions in the projectile
fragmentation regions.
Comparing  the results of
DPMJET--II with full FZIC with the
restricted superposition model
in Tables~\ref{table3} and \ref{table4} we find:
\begin{enumerate}
\item
For He--air collisions, and presumably for all light nuclei, 
the full model
agrees practically to the model without formation zone cascade.
Therefore, we also find only a 20 \%   disagreement with the
restricted superposition model.

\item
We find that, for O--air and Fe--air, 
differences in the spectrum weighted moments of pions and kaons
are up to 50 -- 70\%  with respect to
the restricted 
superposition model.  We understand the reasons for this large
disagreement looking at the changes in the Feynman $x_F$
distributions due to the FZIC as discussed in the last Section.
The same trend is seen in the energy fractions of pions and
kaons, but there the differences are only around 10\%.

\item
The energy fractions of nuclear fragments $K_{n.f.}$ exist only
in the full model. We find however the sum $K_{b-\bar b} +
K_{n.f.}$ of the full model remarkably close to the energy
fractions $K^{sup}_{b-\bar b}$ of the restricted superposition model.
\end{enumerate}

\section{Conclusions and summary}

We have compared Feynman--$x_F$ distributions, spectrum weighted
moments  and energy fractions of pions, kaons  and baryons
minus antibaryons in two versions of the two component
Dual Parton Model. 
One version is the full model with FZIC of the produced hadrons with 
the spectators of the target and projectile nuclei, the second version 
is the DPM (closer to a restricted superposition model)
without this formation zone cascade, nuclear
evaporation and the formation of a residual nucleus.

We have discussed the reasons for deviations expected 
between the DPM and
restricted superposition models.
Nevertheless, we  find a reasonable, however not perfect,
agreement of the moments from the restricted model for
nucleus--nucleus collisions with a restricted superposition model. 

In the full model instead, 
the hadrons produced in the projectile fragmentation region by
the formation zone cascade lead to  significant deviations in
the pion spectrum weighted moments from the restricted 
superposition model. For
Fe--air collisions this difference becomes as large as 70\%.
This effect depends on the changes in the Feynman--$x_F$
distributions due to the cascade in the projectile nucleus. We
have stressed above that there is no direct experimental evidence
for this. Therefore we cannot be completely sure about the
quantitative size of the  predicted effect, but we are convinced that
this effect exists.

These differences are certainly significant
and could also lead to differences when sampling the cosmic
ray cascade with the full DPMJET--II model as compared to the
sampling using the superposition model. 
Larger values for $Z_{\pi}$ and $K_{\pi}$ 
(also for $\pi^0$, where the differences
are similar, and kaons) mean that, 
on the average, the primary cosmic ray
loses a larger fraction of its energy in the first interaction. This 
energy is spent for pion production. 
On the other hand, we expect that less energy
is carried away by the leading particles, so that the development of 
the shower might be different from that resulting from the superposition
model. Depending on the relevance of the first interaction,
this should probably lead to showers 
with a larger muon content and with a smaller depth of the maximum 
development (thus with a smaller electromagnetic size at mountain or 
sea level). 
We might also expect that in the simulation of the full cosmic ray
cascade initiated by primary nuclei, the
full model could well show further deviations
from the superposition model which are not simply contained in the
differences of spectrum weighted moments  and energy fractions
of pions and kaons.
This is another reason to stress the importance of studying the
full model in shower simulations. 

In summary, DPMJET--II including FZIC offers now
the opportunity to perform a true quantitative test of 
the validity (or failure) of the superposition approximation.
A detailed study of the effects of the FZIC model on cosmic ray 
showers as a function of the primary energy and mass will be
the object of a next paper.
\par\noindent

{\bf Acknowledgements}
\par
We thank T.K.Gaisser and P.Lipari for useful discussions.
One of the authors (J.R.) thanks Professor C.Pajares for the
hospitality at Santiago de Compostela and he was supported by the
Direccion General de Politicia Cientifica of Spain.

\par\noindent
%{\bf References}
\par\noindent
%
%-------------- Bibliography -------------------
%\clearpage
%\bibliographystyle{zpc}
%\bibliography{dpm10}

\begin{thebibliography}{10}

\bibitem{HEMAS}
{C.~Forti, H.~Bilokon, B.~d'Ettorre Piazzoli, T.K.~Gaisser, L.~Satta and
  T.~Stanev}:
\newblock Phys. \ Rev. \ D42 (1990) 3668

\bibitem{SIBYLL}
{ R.S.~Fletcher, T.K.~Gaisser, P.~Lipari and T.~Stanev,}:
\newblock Phys.\ Rev.\ D50 (1994) 5710

\bibitem{Capella94a}
A.~Capella, U.~Sukhatme, C.~I. Tan  and J.~Tran Thanh~Van:
\newblock Phys.\ Rep.\ 236 (1994) 227

\bibitem{Ranft94c}
J.~Ranft:
\newblock Phys.\ Rev.\ D51 (1995) 64

\bibitem{DPMBFR94}
{G.~Battistoni, C.~Forti and J.~Ranft}:
\newblock Astroparticle Phys. 3 (1995) 157

\bibitem{dpmcharm}
{G.~Battistoni, C.~Bloise, C.~Forti, M.~Greco, J.~Ranft and A.~Tanzini}:
\newblock Astroparticle Physics {4} (1996) 351

\bibitem{Ranft88a}
J.~Ranft:
\newblock Phys.\ Rev.\ D37 (1988) 1842

\bibitem{Ranft89a}
J.~Ranft:
\newblock Z.\ Phys.\ C43 (1989) 439

\bibitem{Moehring91}
H.-J. M\"ohring and J.~Ranft:
\newblock Z.\ Phys.\ C52 (1991) 643

\bibitem{Ferrari95a}
A.~Ferrari, J.~Ranft, S.~Roesler  and P.~R. Sala:
\newblock Z.\ Phys.\ C70 (1996) 413

\bibitem{Ferrari96a}
A.~Ferrari, J.~Ranft, S.~Roesler  and P.~R. Sala:
\newblock Z.\ Phys.\ C71 (1996) 75

\bibitem{Aurenche92a}
P.~Aurenche, F.~W. Bopp, A.~Capella, J.~Kwiecinski, M.~Maire, J.~Ranft  and
  J.~Tran Thanh~Van:
\newblock Phys.\ Rev.\ D45 (1992) 92

\bibitem{Bopp94a}
F.~W. Bopp, R.~Engel, D.~Pertermann  and J.~Ranft:
\newblock Phys.\ Rev.\ D49 (1994) 3236

\bibitem{Shmakov89}
S.~Y. Shmakov, V.~V. Uzhinskii  and A.~M. Zadoroshny:
\newblock Comp.\ Phys.\ Commun.\ 54 (1989) 125

\bibitem{Sjostrand86}
T.~Sj\"ostrand:
\newblock Comp.\ Phys.\ Commun.\ 39 (1986) 347

\bibitem{Sjostrand87a}
T.~Sj\"ostrand and M.~Bengtsson:
\newblock Comp.\ Phys.\ Commun.\ 43 (1987) 367

\bibitem{Bertini63}
H.~W. Bertini:
\newblock Phys.\ Rev.\ 137 (1963) 1801

\bibitem{Bertini69}
H.~W. Bertini:
\newblock Phys.\ Rev.\ 188 (1969) 1711

\bibitem{Stodolski75}
L.~Stodolski:
\newblock Proceedings on the Colloquium on Multiparticle Reactions, Oxford
  (1975) 577

\bibitem{Landau53}
L.~Landau and I.~Pomeranchuk:
\newblock Dokl. Akad. Nauk SSR 92 (1953) 535,734

\bibitem{Bialas84}
A.~Bialas:
\newblock Z.\ Phys.\ C26 (1984) 301

\bibitem{Haenssgen86}
K.~H\"anssgen and J.~Ranft:
\newblock Comp.\ Phys.\ Commun.\ 39 (1986) 37

\bibitem{Ritchie83}
B.~G. Ritchie:
\newblock Phys.\ Rev.\ C 28 (1983) 926

\bibitem{Albrecht93}
WA80 Collab.:  R.~Albrecht et~al.:
\newblock Z.\ Phys.\ C57 (1993) 37

\bibitem{Albrecht93a}
WA80 Collab.:  R.~Albrecht et~al.:
\newblock Phys.\ Lett.\ B309 (1993) 269

\bibitem{gaistext}
T.~K. Gaisser:
\newblock {\em Cosmic Rays and Particle Physics,}
\newblock Cambridge University Press, 1990

\bibitem{cosmicres1}
{J.W.~Elbert, T.K.~Gaisser and T.~Stanev}:
\newblock Phys. \ Rev. \ D27 (1983) 1448

\bibitem{nim85}
T.~K. Gaisser and T.~Stanev:
\newblock Nuclear Instr. \& Meth. A235 (1985) 183

\bibitem{Glauber59}
R.~Glauber:
\newblock {\em in Lectures in Theoretical Physics, eds. W.E.Brittin 
et al., vol 1 p. 315}
\newblock Interscience New York 1959

\bibitem{Franco68}
V.~Franco:
\newblock Phys.\ Rev. 175 (1968) 1376

\bibitem{Czyz69}
W.~Czyz and L.~Maximon:
\newblock Ann. of Phys. (NY) 52 (1969) 59

\bibitem{Kofoed69}
O.~Kofoed-Hansen:
\newblock Nuovo Cim. 60A (1969) 621

\bibitem{Roesler93}
S.~Roesler, R.~Engel  and J.~Ranft:
\newblock Z.\ Phys.\ C59 (1993) 481

\bibitem{Ranft94b}
J.~Ranft and S.~Roesler:
\newblock Z.\ Phys.\ C62 (1994) 329

\end{thebibliography}

% 
%------- Bibliography ----------
%
%\end{thebibliography}
%
%
%++++++++++++++++++++++++++++++++++++++++++++++++++++++++++++++++++++++
%
% 
 \clearpage

\begin{table}[htb]
\caption{\label{table1}
Spectrum weighted moments of pions and kaons 
according to DPMJET--II in the model
without FZIC. In the last columns we compare 
 with the restricted superposition model.
$E$ is the lab--energy per projectile nucleon.
}
\vskip0.5cm
\medskip
\begin{center}
\renewcommand{\arraystretch}{1.5}
\begin{tabular}{|c|c|c|c|c|c|c|} \hline
Collision &  E (TeV) &  $Z_{\pi}$ & $Z_K$  & $\nu_p$  & 
$\nu_pZ_{\pi}^{p-air}$  & $\nu_pZ_K^{p-air}$ \\
\hline
 p--air & 500  &  0.0640 & 0.0092 & 1 &  &   \\
     & $5\cdot 10^{6}$&  0.0595 & 0.0087 & 1 &  &   \\
\hline
 He--air& 500  &  0.178 & 0.025 & 2.32 & 0.148 & 0.021 \\
      & $5\cdot 10^{6}$&  0.180 & 0.026 & 2.60 & 0.155 & 0.023 \\
\hline
 O--air & 500  &  0.425 & 0.059 & 5.56 & 0.356 & 0.051 \\
     & $5\cdot 10^{6}$& 0.462  & 0.065 & 6.49 & 0.386 & 0.057 \\
\hline
 Fe--air& 500 &  0.912 & 0.124 & 11.61 & 0.743 & 0.107  \\
     & $5\cdot 10^{6}$&  1.020 & 0.128 & 14.46 & 0.860 & 0.126 \\
\hline
\end{tabular}
\end{center}
\end{table}

\begin{table}[htb]
\caption{\label{table2}
Energy fractions of pions , kaons and baryons minus antibaryons
according to DPMJET--II in the model
without FZIC. In the last columns we compare 
 with the restricted superposition model.
$E$ is the lab--energy per projectile nucleon.
}
\vskip0.5cm
\medskip
\begin{center}
\renewcommand{\arraystretch}{1.5}
\begin{tabular}{|c|c|c|c|c|c|c|c|c|} \hline
Collision &  E (TeV) &  $K_{\pi}$ & $K_K$ &$K_{b-\bar b}$  & $\nu_p$  & 
$\nu_pK_{\pi}^{p-air}$  & $\nu_pK_K^{p-air}$ &$\nu_pK_{b-\bar b}^{p-Air}$ \\
\hline
 p--air & 500  &  0.336 & 0.048 &0.268 & 1 &  & &  \\
     & $5\cdot 10^{6}$&  0.361 & 0.052 &0.193 & 1 &  & &   \\
\hline
 He--air& 500  &  0.89 & 0.13 & 0.82 & 2.32 & 0.78 &0.11 &0.62 \\
      & $5\cdot 10^{6}$&  1.01 & 0.14 & 0.76 & 2.60 & 0.94 &0.14 &0.50\\
\hline
 O--air & 500  &  2.02 & 0.28 &1.22 & 5.56 & 1.87 & 0.27 &1.49 \\
     & $5\cdot 10^{6}$& 2.43  & 0.32 &1.24 & 6.49 & 2.34 & 0.34 &1.25\\
\hline
 Fe--air& 500 &  4.05 & 0.54 &2.37 & 11.61 & 3.90 & 0.56 &3.11 \\
     & $5\cdot 10^{6}$&  4.70 & 0.60 &2.35 & 14.46 & 5.22 & 0.75 &2.79 \\
\hline
\end{tabular}
\end{center}
\end{table}

\begin{table}[htb]
\caption{\label{table3}
Spectrum weighted moments of pions and kaons 
according to DPMJET--II in the model
with full FZIC. We give also the values of
$\nu$, $\nu_p$ and $\nu_t$ for the minimum bias Glauber cascade
introduced in Section 2.1. $E$ is the lab--energy per projectile nucleon.
 In the last columns we compare 
 with the restricted superposition model.
}
\vskip0.5cm
\medskip
\begin{center}
\renewcommand{\arraystretch}{1.5}
\begin{tabular}{|c|c|c|c|c|c|c|c|c|} \hline
Collision &  E (TeV)  & $\nu$ & $\nu_p$ & $\nu_t$ & $Z_{\pi}$ & $Z_K$  & 
$\nu_pZ_{\pi}^{p-air}$  & $\nu_pZ_K^{p-air}$ \\
\hline
 p--air & 500  & 2.05 & 1. & 2.05 & 0.0640 & 0.0091 &  &   \\
     & $5\cdot 10^{6}$& 2.73 & 1. & 2.73 & 0.0586 & 0.0085 &  & \\
\hline
 He--air& 500  & 4.44 & 2.32 & 3.41 & 0.182 & 0.026 & 0.148 & 0.021 \\
      & $5\cdot 10^{6}$& 6.58 & 2.60 & 4.37 & 0.175 & 0.025 & 0.152 & 0.022 \\
\hline
 O--air & 500  & 10.34 & 5.55 & 5.32 & 0.499 & 0.065 & 0.355 & 0.051 \\
     & $5\cdot 10^{6}$& 14.35 & 6.22 & 5.81 & 0.522 & 0.065 & 0.364 & 0.053 \\
\hline
 Fe--air& 500 & 21.51 & 11.89 & 6.87 & 1.274 & 0.137 & 0.761 & 0.101 \\
  & $5\cdot 10^{6}$& 32.48 & 13.86 & 7.25 & 1.287 & 0.137 & 0.812 & 0.118 \\
\hline
\end{tabular}
\end{center}
\end{table}

\begin{table}[htb]
\caption{\label{table4}
Energy fractions of pions , kaons , baryons minus antibaryons
and nuclear fragments
according to DPMJET--II in the model
with full FZIC. We give also the values of
 $\nu_p$  for the minimum bias Glauber cascade
introduced in Section 2.1. $E$ is the lab--energy per projectile nucleon.
 In the last columns we compare 
 with the restricted superposition model.
}
\vskip0.5cm
\medskip
\begin{center}
\renewcommand{\arraystretch}{1.5}
%
\vspace*{1cm}
\caption{
Feynman--$x_F$ distribution  of $\pi^+$ 
in p--air collisions at a lab--energy
of 500 TeV.
Plot 1 (large full circles) is for the full model with FZIC 
in the target nucleus; plot 2 (small full circles) is for the model
without FZIC. The target fragmentation region
is at negative Feynman--$x_F$, the projectile fragmentation is at
positive $x_F$. For comparison we give also the same
distribution in p--p collisions (small empty circles). 
\protect\label{pairxfpip1e3n}
}
\end{figure}
 \clearpage

\begin{figure}[thb] \centering
\hspace*{0.25cm}
%\input{heairxfpip1e3n}
% GNUPLOT: LaTeX picture
\setlength{\unitlength}{0.240900pt}
\ifx\plotpoint\undefined\newsavebox{\plotpoint}\fi

\vspace*{1cm}
\caption{
Feynman--$x_F$ distribution of $\pi^+$ 
in He--air collisions at a lab--energy
of 500 TeV per projectile nucleon.
Plot 1 (large full circles) is for the full model with FZIC 
in the target nucleus, and projectile nucleus;
plot 2 (small full circles) is for the model
without FZIC. The target fragmentation region
is at negative Feynman--$x_F$, the projectile fragmentation is at
positive $x_F$. For comparison we give also the same
distribution in p--air collisions with FZIC (small empty circles).
\protect\label{heairxfpip1e3n}
}
\end{figure}
 \clearpage

\begin{figure}[thb] \centering
\hspace*{0.25cm}
%\input{oairxfpip1e3n}
% GNUPLOT: LaTeX picture
\setlength{\unitlength}{0.240900pt}
\ifx\plotpoint\undefined\newsavebox{\plotpoint}\fi

\vspace*{1cm}
\caption{
Feynman--$x_F$ distribution of $\pi^+$ 
in O--air collisions at a lab--energy
of 500 TeV per projectile nucleon.
Plot 1 (large full circles) is for the full model with FZIC 
in the target nucleus, and projectile nucleus;
plot 2 (small full circles) is for the model
without FZIC. The target fragmentation region
is at negative Feynman--$x_F$, the projectile fragmentation is at
positive $x_F$. For comparison we give also the same
distribution in p--air collisions with FZIC (small empty circles).
\protect\label{oairxfpip1e3n}
}
\end{figure}
 \clearpage

\begin{figure}[thb] \centering
\hspace*{0.25cm}
%\input{feairxfpip1e3n}
% GNUPLOT: LaTeX picture
\setlength{\unitlength}{0.240900pt}
\ifx\plotpoint\undefined\newsavebox{\plotpoint}\fi

\vspace*{1cm}
\caption{
Feynman--$x_F$ distribution of $\pi^+$ 
in Fe--air collisions at a lab--energy
of 500 TeV per projectile nucleon.
Plot 1 (large full circles) is for the full model with FZIC 
in the target nucleus, and projectile nucleus;
plot 2 (small full circles) is for the model
without FZIC. The target fragmentation region
is at negative Feynman--$x_F$, the projectile fragmentation is at
positive $x_F$. For comparison we give also the same
distribution in p--air collisions with FZIC (small empty circles).
\protect\label{feairxfpip1e3n}
}
\end{figure}
 \clearpage

\begin{figure}[thb] \centering
\hspace*{0.25cm}
%\input{feairxfpip1e3n}
% GNUPLOT: LaTeX picture
\setlength{\unitlength}{0.240900pt}
\ifx\plotpoint\undefined\newsavebox{\plotpoint}\fi
\sbox{\plotpoint}{\rule[-0.200pt]{0.400pt}{0.400pt}}%
\begin{picture}(1653,1653)(0,0)
\font\gnuplot=cmr10 at 10pt
\gnuplot
\sbox{\plotpoint}{\rule[-0.200pt]{0.400pt}{0.400pt}}%
\put(220.0,113.0){\rule[-0.200pt]{4.818pt}{0.400pt}}
\put(198,113){\makebox(0,0)[r]{$0$}}
\put(1569.0,113.0){\rule[-0.200pt]{4.818pt}{0.400pt}}
\put(220.0,315.0){\rule[-0.200pt]{4.818pt}{0.400pt}}
\put(198,315){\makebox(0,0)[r]{$0.2$}}
\put(1569.0,315.0){\rule[-0.200pt]{4.818pt}{0.400pt}}
\put(220.0,518.0){\rule[-0.200pt]{4.818pt}{0.400pt}}
\put(198,518){\makebox(0,0)[r]{$0.4$}}
\put(1569.0,518.0){\rule[-0.200pt]{4.818pt}{0.400pt}}
\put(220.0,720.0){\rule[-0.200pt]{4.818pt}{0.400pt}}
\put(198,720){\makebox(0,0)[r]{$0.6$}}
\put(1569.0,720.0){\rule[-0.200pt]{4.818pt}{0.400pt}}
\put(220.0,922.0){\rule[-0.200pt]{4.818pt}{0.400pt}}
\put(198,922){\makebox(0,0)[r]{$0.8$}}
\put(1569.0,922.0){\rule[-0.200pt]{4.818pt}{0.400pt}}
\put(220.0,1124.0){\rule[-0.200pt]{4.818pt}{0.400pt}}
\put(198,1124){\makebox(0,0)[r]{$1$}}
\put(1569.0,1124.0){\rule[-0.200pt]{4.818pt}{0.400pt}}
\put(220.0,1327.0){\rule[-0.200pt]{4.818pt}{0.400pt}}
\put(198,1327){\makebox(0,0)[r]{$1.2$}}
\put(1569.0,1327.0){\rule[-0.200pt]{4.818pt}{0.400pt}}
\put(220.0,1529.0){\rule[-0.200pt]{4.818pt}{0.400pt}}
\put(198,1529){\makebox(0,0)[r]{$1.4$}}
\put(1569.0,1529.0){\rule[-0.200pt]{4.818pt}{0.400pt}}
\put(220.0,113.0){\rule[-0.200pt]{0.400pt}{4.818pt}}
\put(220,68){\makebox(0,0){$0$}}
\put(220.0,1610.0){\rule[-0.200pt]{0.400pt}{4.818pt}}
\put(403.0,113.0){\rule[-0.200pt]{0.400pt}{4.818pt}}
\put(403,68){\makebox(0,0){$2$}}
\put(403.0,1610.0){\rule[-0.200pt]{0.400pt}{4.818pt}}
\put(585.0,113.0){\rule[-0.200pt]{0.400pt}{4.818pt}}
\put(585,68){\makebox(0,0){$4$}}
\put(585.0,1610.0){\rule[-0.200pt]{0.400pt}{4.818pt}}
\put(768.0,113.0){\rule[-0.200pt]{0.400pt}{4.818pt}}
\put(768,68){\makebox(0,0){$6$}}
\put(768.0,1610.0){\rule[-0.200pt]{0.400pt}{4.818pt}}
\put(950.0,113.0){\rule[-0.200pt]{0.400pt}{4.818pt}}
\put(950,68){\makebox(0,0){$8$}}
\put(950.0,1610.0){\rule[-0.200pt]{0.400pt}{4.818pt}}
\put(1133.0,113.0){\rule[-0.200pt]{0.400pt}{4.818pt}}
\put(1133,68){\makebox(0,0){$10$}}
\put(1133.0,1610.0){\rule[-0.200pt]{0.400pt}{4.818pt}}
\put(1315.0,113.0){\rule[-0.200pt]{0.400pt}{4.818pt}}
\put(1315,68){\makebox(0,0){$12$}}
\put(1315.0,1610.0){\rule[-0.200pt]{0.400pt}{4.818pt}}
\put(1498.0,113.0){\rule[-0.200pt]{0.400pt}{4.818pt}}
\put(1498,68){\makebox(0,0){$14$}}
\put(1498.0,1610.0){\rule[-0.200pt]{0.400pt}{4.818pt}}
\put(220.0,113.0){\rule[-0.200pt]{329.792pt}{0.400pt}}
\put(1589.0,113.0){\rule[-0.200pt]{0.400pt}{365.445pt}}
\put(220.0,1630.0){\rule[-0.200pt]{329.792pt}{0.400pt}}
\put(45,871){\makebox(0,0){$\large \Gamma(b)$}}
\put(904,23){\makebox(0,0){$\large b$~~~~~(fm)}}
\put(220.0,113.0){\rule[-0.200pt]{0.400pt}{365.445pt}}
\put(1459,1565){\makebox(0,0)[r]{Fe-N}}
\put(1481.0,1565.0){\rule[-0.200pt]{15.899pt}{0.400pt}}
\put(234,1124){\usebox{\plotpoint}}
\put(699,1122.67){\rule{3.373pt}{0.400pt}}
\multiput(699.00,1123.17)(7.000,-1.000){2}{\rule{1.686pt}{0.400pt}}
\put(713,1121.67){\rule{3.373pt}{0.400pt}}
\multiput(713.00,1122.17)(7.000,-1.000){2}{\rule{1.686pt}{0.400pt}}
\put(727,1120.67){\rule{3.132pt}{0.400pt}}
\multiput(727.00,1121.17)(6.500,-1.000){2}{\rule{1.566pt}{0.400pt}}
\put(740,1119.17){\rule{2.900pt}{0.400pt}}
\multiput(740.00,1120.17)(7.981,-2.000){2}{\rule{1.450pt}{0.400pt}}
\multiput(754.00,1117.94)(1.943,-0.468){5}{\rule{1.500pt}{0.113pt}}
\multiput(754.00,1118.17)(10.887,-4.000){2}{\rule{0.750pt}{0.400pt}}
\multiput(768.00,1113.93)(1.123,-0.482){9}{\rule{0.967pt}{0.116pt}}
\multiput(768.00,1114.17)(10.994,-6.000){2}{\rule{0.483pt}{0.400pt}}
\multiput(781.00,1107.93)(1.026,-0.485){11}{\rule{0.900pt}{0.117pt}}
\multiput(781.00,1108.17)(12.132,-7.000){2}{\rule{0.450pt}{0.400pt}}
\multiput(795.00,1100.92)(0.637,-0.492){19}{\rule{0.609pt}{0.118pt}}
\multiput(795.00,1101.17)(12.736,-11.000){2}{\rule{0.305pt}{0.400pt}}
\multiput(809.00,1089.92)(0.497,-0.493){23}{\rule{0.500pt}{0.119pt}}
\multiput(809.00,1090.17)(11.962,-13.000){2}{\rule{0.250pt}{0.400pt}}
\multiput(822.58,1075.45)(0.494,-0.644){25}{\rule{0.119pt}{0.614pt}}
\multiput(821.17,1076.73)(14.000,-16.725){2}{\rule{0.400pt}{0.307pt}}
\multiput(836.58,1056.98)(0.494,-0.791){25}{\rule{0.119pt}{0.729pt}}
\multiput(835.17,1058.49)(14.000,-20.488){2}{\rule{0.400pt}{0.364pt}}
\multiput(850.58,1034.14)(0.493,-1.052){23}{\rule{0.119pt}{0.931pt}}
\multiput(849.17,1036.07)(13.000,-25.068){2}{\rule{0.400pt}{0.465pt}}
\multiput(863.58,1006.67)(0.494,-1.195){25}{\rule{0.119pt}{1.043pt}}
\multiput(862.17,1008.84)(14.000,-30.835){2}{\rule{0.400pt}{0.521pt}}
\multiput(877.58,972.96)(0.494,-1.415){25}{\rule{0.119pt}{1.214pt}}
\multiput(876.17,975.48)(14.000,-36.480){2}{\rule{0.400pt}{0.607pt}}
\multiput(891.58,933.48)(0.494,-1.562){25}{\rule{0.119pt}{1.329pt}}
\multiput(890.17,936.24)(14.000,-40.242){2}{\rule{0.400pt}{0.664pt}}
\multiput(905.58,889.45)(0.493,-1.884){23}{\rule{0.119pt}{1.577pt}}
\multiput(904.17,892.73)(13.000,-44.727){2}{\rule{0.400pt}{0.788pt}}
\multiput(918.58,841.42)(0.494,-1.892){25}{\rule{0.119pt}{1.586pt}}
\multiput(917.17,844.71)(14.000,-48.709){2}{\rule{0.400pt}{0.793pt}}
\multiput(932.58,789.18)(0.494,-1.966){25}{\rule{0.119pt}{1.643pt}}
\multiput(931.17,792.59)(14.000,-50.590){2}{\rule{0.400pt}{0.821pt}}
\multiput(946.58,734.43)(0.493,-2.201){23}{\rule{0.119pt}{1.823pt}}
\multiput(945.17,738.22)(13.000,-52.216){2}{\rule{0.400pt}{0.912pt}}
\multiput(959.58,678.94)(0.494,-2.039){25}{\rule{0.119pt}{1.700pt}}
\multiput(958.17,682.47)(14.000,-52.472){2}{\rule{0.400pt}{0.850pt}}
\multiput(973.58,622.94)(0.494,-2.039){25}{\rule{0.119pt}{1.700pt}}
\multiput(972.17,626.47)(14.000,-52.472){2}{\rule{0.400pt}{0.850pt}}
\multiput(987.58,566.56)(0.493,-2.162){23}{\rule{0.119pt}{1.792pt}}
\multiput(986.17,570.28)(13.000,-51.280){2}{\rule{0.400pt}{0.896pt}}
\multiput(1000.58,512.42)(0.494,-1.892){25}{\rule{0.119pt}{1.586pt}}
\multiput(999.17,515.71)(14.000,-48.709){2}{\rule{0.400pt}{0.793pt}}
\multiput(1014.58,460.77)(0.494,-1.782){25}{\rule{0.119pt}{1.500pt}}
\multiput(1013.17,463.89)(14.000,-45.887){2}{\rule{0.400pt}{0.750pt}}
\multiput(1028.58,411.71)(0.493,-1.805){23}{\rule{0.119pt}{1.515pt}}
\multiput(1027.17,414.85)(13.000,-42.855){2}{\rule{0.400pt}{0.758pt}}
\multiput(1041.58,366.72)(0.494,-1.488){25}{\rule{0.119pt}{1.271pt}}
\multiput(1040.17,369.36)(14.000,-38.361){2}{\rule{0.400pt}{0.636pt}}
\multiput(1055.58,326.32)(0.494,-1.305){25}{\rule{0.119pt}{1.129pt}}
\multiput(1054.17,328.66)(14.000,-33.658){2}{\rule{0.400pt}{0.564pt}}
\multiput(1069.58,290.50)(0.493,-1.250){23}{\rule{0.119pt}{1.085pt}}
\multiput(1068.17,292.75)(13.000,-29.749){2}{\rule{0.400pt}{0.542pt}}
\multiput(1082.58,259.38)(0.494,-0.974){25}{\rule{0.119pt}{0.871pt}}
\multiput(1081.17,261.19)(14.000,-25.191){2}{\rule{0.400pt}{0.436pt}}
\multiput(1096.58,232.86)(0.494,-0.827){25}{\rule{0.119pt}{0.757pt}}
\multiput(1095.17,234.43)(14.000,-21.429){2}{\rule{0.400pt}{0.379pt}}
\multiput(1110.58,210.21)(0.494,-0.717){25}{\rule{0.119pt}{0.671pt}}
\multiput(1109.17,211.61)(14.000,-18.606){2}{\rule{0.400pt}{0.336pt}}
\multiput(1124.58,190.54)(0.493,-0.616){23}{\rule{0.119pt}{0.592pt}}
\multiput(1123.17,191.77)(13.000,-14.771){2}{\rule{0.400pt}{0.296pt}}
\multiput(1137.00,175.92)(0.497,-0.494){25}{\rule{0.500pt}{0.119pt}}
\multiput(1137.00,176.17)(12.962,-14.000){2}{\rule{0.250pt}{0.400pt}}
\multiput(1151.00,161.92)(0.637,-0.492){19}{\rule{0.609pt}{0.118pt}}
\multiput(1151.00,162.17)(12.736,-11.000){2}{\rule{0.305pt}{0.400pt}}
\multiput(1165.00,150.93)(0.728,-0.489){15}{\rule{0.678pt}{0.118pt}}
\multiput(1165.00,151.17)(11.593,-9.000){2}{\rule{0.339pt}{0.400pt}}
\multiput(1178.00,141.93)(1.026,-0.485){11}{\rule{0.900pt}{0.117pt}}
\multiput(1178.00,142.17)(12.132,-7.000){2}{\rule{0.450pt}{0.400pt}}
\multiput(1192.00,134.93)(1.214,-0.482){9}{\rule{1.033pt}{0.116pt}}
\multiput(1192.00,135.17)(11.855,-6.000){2}{\rule{0.517pt}{0.400pt}}
\multiput(1206.00,128.94)(1.797,-0.468){5}{\rule{1.400pt}{0.113pt}}
\multiput(1206.00,129.17)(10.094,-4.000){2}{\rule{0.700pt}{0.400pt}}
\multiput(1219.00,124.94)(1.943,-0.468){5}{\rule{1.500pt}{0.113pt}}
\multiput(1219.00,125.17)(10.887,-4.000){2}{\rule{0.750pt}{0.400pt}}
\put(1233,120.17){\rule{2.900pt}{0.400pt}}
\multiput(1233.00,121.17)(7.981,-2.000){2}{\rule{1.450pt}{0.400pt}}
\put(1247,118.17){\rule{2.700pt}{0.400pt}}
\multiput(1247.00,119.17)(7.396,-2.000){2}{\rule{1.350pt}{0.400pt}}
\put(1260,116.17){\rule{2.900pt}{0.400pt}}
\multiput(1260.00,117.17)(7.981,-2.000){2}{\rule{1.450pt}{0.400pt}}
\put(1274,114.67){\rule{3.373pt}{0.400pt}}
\multiput(1274.00,115.17)(7.000,-1.000){2}{\rule{1.686pt}{0.400pt}}
\put(1288,113.67){\rule{3.373pt}{0.400pt}}
\multiput(1288.00,114.17)(7.000,-1.000){2}{\rule{1.686pt}{0.400pt}}
\put(234.0,1124.0){\rule[-0.200pt]{112.018pt}{0.400pt}}
\put(1329,112.67){\rule{3.373pt}{0.400pt}}
\multiput(1329.00,113.17)(7.000,-1.000){2}{\rule{1.686pt}{0.400pt}}
\put(1302.0,114.0){\rule[-0.200pt]{6.504pt}{0.400pt}}
\put(1343.0,113.0){\rule[-0.200pt]{59.261pt}{0.400pt}}
\sbox{\plotpoint}{\rule[-0.500pt]{1.000pt}{1.000pt}}%
\put(1459,1520){\makebox(0,0)[r]{He-N}}
\multiput(1481,1520)(20.756,0.000){4}{\usebox{\plotpoint}}
\put(1547,1520){\usebox{\plotpoint}}
\put(234,1124){\usebox{\plotpoint}}
\put(234.00,1124.00){\usebox{\plotpoint}}
\put(254.76,1124.00){\usebox{\plotpoint}}
\multiput(261,1124)(20.756,0.000){0}{\usebox{\plotpoint}}
\put(275.51,1123.96){\usebox{\plotpoint}}
\put(296.23,1123.00){\usebox{\plotpoint}}
\multiput(302,1123)(20.756,0.000){0}{\usebox{\plotpoint}}
\put(316.98,1122.93){\usebox{\plotpoint}}
\put(337.68,1121.41){\usebox{\plotpoint}}
\multiput(343,1121)(20.703,-1.479){0}{\usebox{\plotpoint}}
\put(358.38,1119.90){\usebox{\plotpoint}}
\put(379.01,1117.77){\usebox{\plotpoint}}
\multiput(384,1117)(20.547,-2.935){0}{\usebox{\plotpoint}}
\put(399.50,1114.57){\usebox{\plotpoint}}
\put(419.24,1108.21){\usebox{\plotpoint}}
\put(438.05,1099.48){\usebox{\plotpoint}}
\multiput(439,1099)(18.021,-10.298){0}{\usebox{\plotpoint}}
\put(455.83,1088.83){\usebox{\plotpoint}}
\put(471.80,1075.61){\usebox{\plotpoint}}
\put(486.53,1061.00){\usebox{\plotpoint}}
\put(499.96,1045.21){\usebox{\plotpoint}}
\put(512.08,1028.38){\usebox{\plotpoint}}
\multiput(521,1015)(10.141,-18.109){2}{\usebox{\plotpoint}}
\put(543.07,974.43){\usebox{\plotpoint}}
\multiput(549,963)(7.812,-19.229){2}{\usebox{\plotpoint}}
\multiput(562,931)(7.903,-19.192){2}{\usebox{\plotpoint}}
\multiput(576,897)(7.175,-19.476){2}{\usebox{\plotpoint}}
\multiput(590,859)(6.273,-19.785){2}{\usebox{\plotpoint}}
\multiput(603,818)(6.426,-19.736){2}{\usebox{\plotpoint}}
\multiput(617,775)(6.166,-19.819){2}{\usebox{\plotpoint}}
\multiput(631,730)(5.533,-20.004){2}{\usebox{\plotpoint}}
\multiput(644,683)(5.925,-19.892){3}{\usebox{\plotpoint}}
\multiput(658,636)(5.812,-19.925){2}{\usebox{\plotpoint}}
\multiput(672,588)(5.533,-20.004){2}{\usebox{\plotpoint}}
\multiput(685,541)(6.043,-19.856){3}{\usebox{\plotpoint}}
\multiput(699,495)(6.426,-19.736){2}{\usebox{\plotpoint}}
\multiput(713,452)(6.563,-19.690){2}{\usebox{\plotpoint}}
\multiput(727,410)(6.563,-19.690){2}{\usebox{\plotpoint}}
\multiput(740,371)(7.708,-19.271){2}{\usebox{\plotpoint}}
\multiput(754,336)(8.106,-19.107){2}{\usebox{\plotpoint}}
\put(775.00,286.85){\usebox{\plotpoint}}
\multiput(781,273)(9.840,-18.275){2}{\usebox{\plotpoint}}
\put(804.17,231.93){\usebox{\plotpoint}}
\put(815.48,214.53){\usebox{\plotpoint}}
\put(827.85,197.90){\usebox{\plotpoint}}
\put(841.61,182.39){\usebox{\plotpoint}}
\put(856.53,167.97){\usebox{\plotpoint}}
\put(872.73,155.05){\usebox{\plotpoint}}
\put(890.46,144.31){\usebox{\plotpoint}}
\multiput(891,144)(19.077,-8.176){0}{\usebox{\plotpoint}}
\put(909.46,135.94){\usebox{\plotpoint}}
\put(928.91,128.88){\usebox{\plotpoint}}
\multiput(932,128)(20.295,-4.349){0}{\usebox{\plotpoint}}
\put(949.14,124.28){\usebox{\plotpoint}}
\put(969.53,120.50){\usebox{\plotpoint}}
\multiput(973,120)(20.547,-2.935){0}{\usebox{\plotpoint}}
\put(990.10,117.76){\usebox{\plotpoint}}
\put(1010.80,116.23){\usebox{\plotpoint}}
\multiput(1014,116)(20.703,-1.479){0}{\usebox{\plotpoint}}
\put(1031.51,115.00){\usebox{\plotpoint}}
\put(1052.24,114.20){\usebox{\plotpoint}}
\multiput(1055,114)(20.756,0.000){0}{\usebox{\plotpoint}}
\put(1072.98,114.00){\usebox{\plotpoint}}
\put(1093.74,114.00){\usebox{\plotpoint}}
\multiput(1096,114)(20.703,-1.479){0}{\usebox{\plotpoint}}
\put(1114.46,113.00){\usebox{\plotpoint}}
\put(1135.21,113.00){\usebox{\plotpoint}}
\multiput(1137,113)(20.756,0.000){0}{\usebox{\plotpoint}}
\put(1155.97,113.00){\usebox{\plotpoint}}
\put(1176.73,113.00){\usebox{\plotpoint}}
\multiput(1178,113)(20.756,0.000){0}{\usebox{\plotpoint}}
\put(1197.48,113.00){\usebox{\plotpoint}}
\put(1218.24,113.00){\usebox{\plotpoint}}
\multiput(1219,113)(20.756,0.000){0}{\usebox{\plotpoint}}
\put(1238.99,113.00){\usebox{\plotpoint}}
\put(1259.75,113.00){\usebox{\plotpoint}}
\multiput(1260,113)(20.756,0.000){0}{\usebox{\plotpoint}}
\put(1280.50,113.00){\usebox{\plotpoint}}
\put(1301.26,113.00){\usebox{\plotpoint}}
\multiput(1302,113)(20.756,0.000){0}{\usebox{\plotpoint}}
\put(1322.01,113.00){\usebox{\plotpoint}}
\put(1342.77,113.00){\usebox{\plotpoint}}
\multiput(1343,113)(20.756,0.000){0}{\usebox{\plotpoint}}
\put(1363.53,113.00){\usebox{\plotpoint}}
\multiput(1370,113)(20.756,0.000){0}{\usebox{\plotpoint}}
\put(1384.28,113.00){\usebox{\plotpoint}}
\put(1405.04,113.00){\usebox{\plotpoint}}
\multiput(1411,113)(20.756,0.000){0}{\usebox{\plotpoint}}
\put(1425,113){\usebox{\plotpoint}}
\sbox{\plotpoint}{\rule[-0.400pt]{0.800pt}{0.800pt}}%
\put(1459,1475){\makebox(0,0)[r]{p-N}}
\put(1481.0,1475.0){\rule[-0.400pt]{15.899pt}{0.800pt}}
\put(234,1083){\usebox{\plotpoint}}
\put(247,1080.84){\rule{3.373pt}{0.800pt}}
\multiput(247.00,1081.34)(7.000,-1.000){2}{\rule{1.686pt}{0.800pt}}
\put(261,1079.34){\rule{3.373pt}{0.800pt}}
\multiput(261.00,1080.34)(7.000,-2.000){2}{\rule{1.686pt}{0.800pt}}
\put(275,1076.84){\rule{3.132pt}{0.800pt}}
\multiput(275.00,1078.34)(6.500,-3.000){2}{\rule{1.566pt}{0.800pt}}
\put(288,1073.84){\rule{3.373pt}{0.800pt}}
\multiput(288.00,1075.34)(7.000,-3.000){2}{\rule{1.686pt}{0.800pt}}
\multiput(302.00,1072.06)(1.936,-0.560){3}{\rule{2.440pt}{0.135pt}}
\multiput(302.00,1072.34)(8.936,-5.000){2}{\rule{1.220pt}{0.800pt}}
\multiput(316.00,1067.07)(1.355,-0.536){5}{\rule{2.067pt}{0.129pt}}
\multiput(316.00,1067.34)(9.711,-6.000){2}{\rule{1.033pt}{0.800pt}}
\multiput(330.00,1061.08)(1.000,-0.526){7}{\rule{1.686pt}{0.127pt}}
\multiput(330.00,1061.34)(9.501,-7.000){2}{\rule{0.843pt}{0.800pt}}
\multiput(343.00,1054.08)(0.800,-0.516){11}{\rule{1.444pt}{0.124pt}}
\multiput(343.00,1054.34)(11.002,-9.000){2}{\rule{0.722pt}{0.800pt}}
\multiput(357.00,1045.08)(0.639,-0.512){15}{\rule{1.218pt}{0.123pt}}
\multiput(357.00,1045.34)(11.472,-11.000){2}{\rule{0.609pt}{0.800pt}}
\multiput(372.41,1031.59)(0.509,-0.533){19}{\rule{0.123pt}{1.062pt}}
\multiput(369.34,1033.80)(13.000,-11.797){2}{\rule{0.800pt}{0.531pt}}
\multiput(385.41,1017.14)(0.509,-0.607){21}{\rule{0.123pt}{1.171pt}}
\multiput(382.34,1019.57)(14.000,-14.569){2}{\rule{0.800pt}{0.586pt}}
\multiput(399.41,999.43)(0.509,-0.721){21}{\rule{0.123pt}{1.343pt}}
\multiput(396.34,1002.21)(14.000,-17.213){2}{\rule{0.800pt}{0.671pt}}
\multiput(413.41,978.04)(0.509,-0.947){19}{\rule{0.123pt}{1.677pt}}
\multiput(410.34,981.52)(13.000,-20.519){2}{\rule{0.800pt}{0.838pt}}
\multiput(426.41,953.53)(0.509,-1.026){21}{\rule{0.123pt}{1.800pt}}
\multiput(423.34,957.26)(14.000,-24.264){2}{\rule{0.800pt}{0.900pt}}
\multiput(440.41,924.34)(0.509,-1.217){21}{\rule{0.123pt}{2.086pt}}
\multiput(437.34,928.67)(14.000,-28.671){2}{\rule{0.800pt}{1.043pt}}
\multiput(454.41,889.97)(0.509,-1.443){19}{\rule{0.123pt}{2.415pt}}
\multiput(451.34,894.99)(13.000,-30.987){2}{\rule{0.800pt}{1.208pt}}
\multiput(467.41,853.21)(0.509,-1.560){21}{\rule{0.123pt}{2.600pt}}
\multiput(464.34,858.60)(14.000,-36.604){2}{\rule{0.800pt}{1.300pt}}
\multiput(481.41,810.50)(0.509,-1.675){21}{\rule{0.123pt}{2.771pt}}
\multiput(478.34,816.25)(14.000,-39.248){2}{\rule{0.800pt}{1.386pt}}
\multiput(495.41,763.91)(0.509,-1.939){19}{\rule{0.123pt}{3.154pt}}
\multiput(492.34,770.45)(13.000,-41.454){2}{\rule{0.800pt}{1.577pt}}
\multiput(508.41,716.07)(0.509,-1.904){21}{\rule{0.123pt}{3.114pt}}
\multiput(505.34,722.54)(14.000,-44.536){2}{\rule{0.800pt}{1.557pt}}
\multiput(522.41,664.60)(0.509,-1.980){21}{\rule{0.123pt}{3.229pt}}
\multiput(519.34,671.30)(14.000,-46.299){2}{\rule{0.800pt}{1.614pt}}
\multiput(536.41,611.60)(0.509,-1.980){21}{\rule{0.123pt}{3.229pt}}
\multiput(533.34,618.30)(14.000,-46.299){2}{\rule{0.800pt}{1.614pt}}
\multiput(550.41,557.63)(0.509,-2.146){19}{\rule{0.123pt}{3.462pt}}
\multiput(547.34,564.82)(13.000,-45.815){2}{\rule{0.800pt}{1.731pt}}
\multiput(563.41,506.07)(0.509,-1.904){21}{\rule{0.123pt}{3.114pt}}
\multiput(560.34,512.54)(14.000,-44.536){2}{\rule{0.800pt}{1.557pt}}
\multiput(577.41,455.55)(0.509,-1.827){21}{\rule{0.123pt}{3.000pt}}
\multiput(574.34,461.77)(14.000,-42.773){2}{\rule{0.800pt}{1.500pt}}
\multiput(591.41,406.67)(0.509,-1.815){19}{\rule{0.123pt}{2.969pt}}
\multiput(588.34,412.84)(13.000,-38.837){2}{\rule{0.800pt}{1.485pt}}
\multiput(604.41,363.21)(0.509,-1.560){21}{\rule{0.123pt}{2.600pt}}
\multiput(601.34,368.60)(14.000,-36.604){2}{\rule{0.800pt}{1.300pt}}
\multiput(618.41,322.16)(0.509,-1.408){21}{\rule{0.123pt}{2.371pt}}
\multiput(615.34,327.08)(14.000,-33.078){2}{\rule{0.800pt}{1.186pt}}
\multiput(632.41,284.74)(0.509,-1.319){19}{\rule{0.123pt}{2.231pt}}
\multiput(629.34,289.37)(13.000,-28.370){2}{\rule{0.800pt}{1.115pt}}
\multiput(645.41,253.53)(0.509,-1.026){21}{\rule{0.123pt}{1.800pt}}
\multiput(642.34,257.26)(14.000,-24.264){2}{\rule{0.800pt}{0.900pt}}
\multiput(659.41,226.48)(0.509,-0.874){21}{\rule{0.123pt}{1.571pt}}
\multiput(656.34,229.74)(14.000,-20.738){2}{\rule{0.800pt}{0.786pt}}
\multiput(673.41,202.81)(0.509,-0.823){19}{\rule{0.123pt}{1.492pt}}
\multiput(670.34,205.90)(13.000,-17.903){2}{\rule{0.800pt}{0.746pt}}
\multiput(686.41,183.14)(0.509,-0.607){21}{\rule{0.123pt}{1.171pt}}
\multiput(683.34,185.57)(14.000,-14.569){2}{\rule{0.800pt}{0.586pt}}
\multiput(699.00,169.08)(0.533,-0.509){19}{\rule{1.062pt}{0.123pt}}
\multiput(699.00,169.34)(11.797,-13.000){2}{\rule{0.531pt}{0.800pt}}
\multiput(713.00,156.08)(0.639,-0.512){15}{\rule{1.218pt}{0.123pt}}
\multiput(713.00,156.34)(11.472,-11.000){2}{\rule{0.609pt}{0.800pt}}
\multiput(727.00,145.08)(0.737,-0.516){11}{\rule{1.356pt}{0.124pt}}
\multiput(727.00,145.34)(10.186,-9.000){2}{\rule{0.678pt}{0.800pt}}
\multiput(740.00,136.08)(1.088,-0.526){7}{\rule{1.800pt}{0.127pt}}
\multiput(740.00,136.34)(10.264,-7.000){2}{\rule{0.900pt}{0.800pt}}
\multiput(754.00,129.06)(1.936,-0.560){3}{\rule{2.440pt}{0.135pt}}
\multiput(754.00,129.34)(8.936,-5.000){2}{\rule{1.220pt}{0.800pt}}
\put(768,122.34){\rule{2.800pt}{0.800pt}}
\multiput(768.00,124.34)(7.188,-4.000){2}{\rule{1.400pt}{0.800pt}}
\put(781,118.84){\rule{3.373pt}{0.800pt}}
\multiput(781.00,120.34)(7.000,-3.000){2}{\rule{1.686pt}{0.800pt}}
\put(795,116.34){\rule{3.373pt}{0.800pt}}
\multiput(795.00,117.34)(7.000,-2.000){2}{\rule{1.686pt}{0.800pt}}
\put(809,114.84){\rule{3.132pt}{0.800pt}}
\multiput(809.00,115.34)(6.500,-1.000){2}{\rule{1.566pt}{0.800pt}}
\put(822,113.84){\rule{3.373pt}{0.800pt}}
\multiput(822.00,114.34)(7.000,-1.000){2}{\rule{1.686pt}{0.800pt}}
\put(836,112.84){\rule{3.373pt}{0.800pt}}
\multiput(836.00,113.34)(7.000,-1.000){2}{\rule{1.686pt}{0.800pt}}
\put(234.0,1083.0){\rule[-0.400pt]{3.132pt}{0.800pt}}
\put(863,111.84){\rule{3.373pt}{0.800pt}}
\multiput(863.00,112.34)(7.000,-1.000){2}{\rule{1.686pt}{0.800pt}}
\put(850.0,114.0){\rule[-0.400pt]{3.132pt}{0.800pt}}
\put(877.0,113.0){\rule[-0.400pt]{49.384pt}{0.800pt}}
\end{picture}
\vspace*{1cm}
\caption{
Glauber profile functions $\Gamma(b)$ for Fe--nitrogen,
He--nitrogen and p--nitrogen collisions at 500 TeV.
\protect\label{profilef}
}
\end{figure}
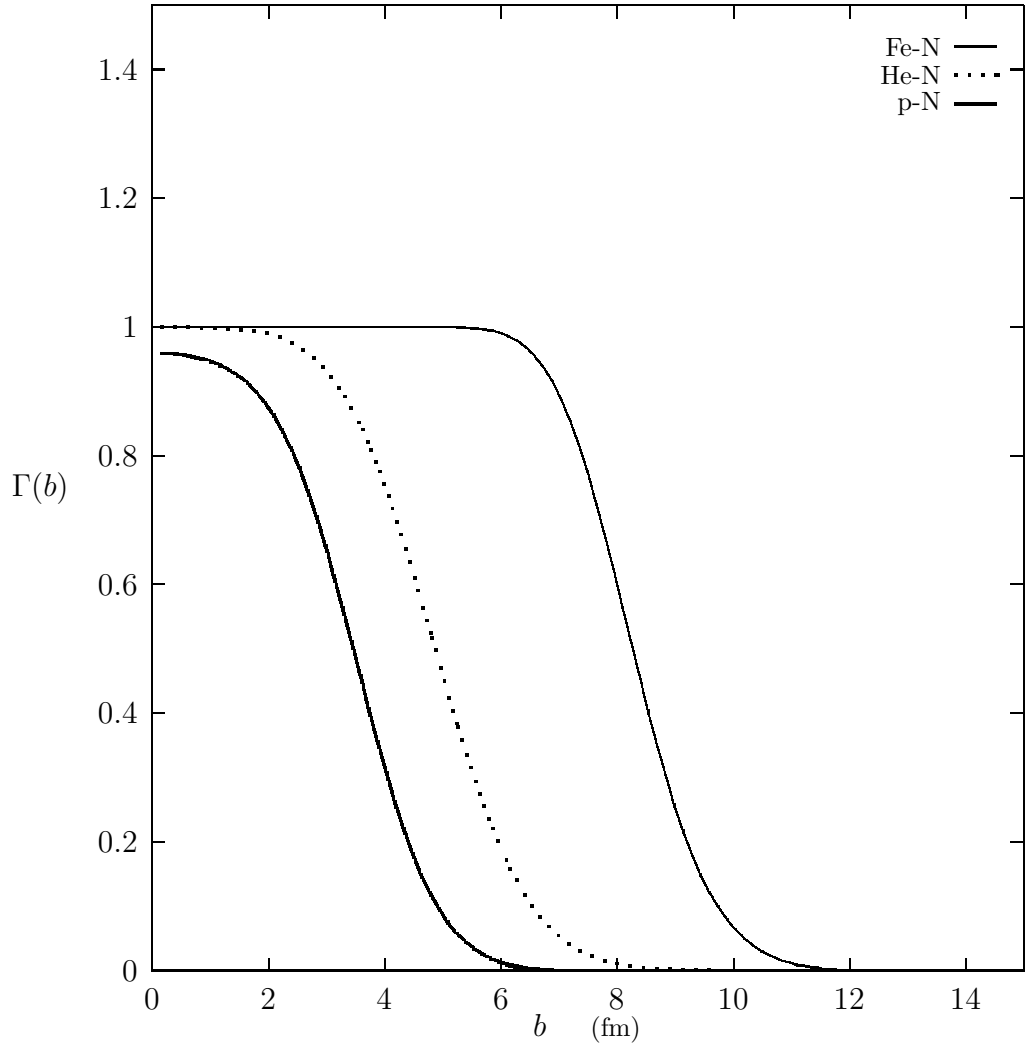
 \clearpage

\end{document}